\documentclass[11pt]{article}
\usepackage{cite}
\usepackage{amsmath,amsfonts,amssymb}
\pdfoutput=1
\usepackage[small,bf,hang]{caption}
\usepackage{slashed}
\input epsf.sty
\usepackage{epsfig}
\usepackage[titletoc,toc]{appendix}
\usepackage{xcolor}

\def\hybrid{
        \topmargin -20pt
        \oddsidemargin 0pt
        \headheight 0pt \headsep 0pt
        \textwidth 6.55in 
        \textheight 9.5in 
        \marginparwidth .875in
        \parskip 5pt plus 1pt \jot = 1.5ex}

\hybrid

 \csname
@addtoreset\endcsname{equation}{section}

\def\moth{\mathsurround=0pt}
\newdimen\zo \zo=0pt

\def\tick{\leaders\hrule height 0.5ex depth 0pt \hskip 0.5pt}
\def\upboxfill{$\moth \setbox\zo\hbox{\tick}%
  \hskip 3pt\hbox to 0pt{$\tick$\hss}\hrulefill \hbox to 7.5pt{$\tick$\hss}$}

\def\dtick{\leaders\hrule height .34pt depth 0.5ex \hskip 0.5pt}
\def\downboxfill{$\moth \setbox\zo\hbox{\dtick}%
  \hskip 2pt\hbox to 0pt{$\dtick$\hss}\hrulefill \hbox to 2pt{$\dtick$\hss}$}

\def\id{{\mathbb I}}

\def\bec{\begin{center}}
\def\ec{\end{center}}

\def\be{\begin{equation}}
\def\ee{\end{equation}}
\def\bea{\begin{eqnarray}}
\def\eea{\end{eqnarray}}
\def\ba{\begin{array}}
\def\ea{\end{array}}

\usepackage{hyperref}

\begin{document}

\begin{titlepage}
	
	\rightline{\tt MIT-CTP-5810}
	\hfill \today
	\begin{center}
		\vskip 0.5cm
		
		{\Large \bf {Boundary terms in string field theory}
		}
		
		\vskip 0.5cm
		
		\vskip 1.0cm
		{\large {Atakan Hilmi Fırat$^{1}$ and Raji Ashenafi Mamade$^{2}$}}
		
		\vskip 0.5cm
		
		{\em  \hskip -.1truecm
			$^{1}$
			Center for Quantum Mathematics and Physics (QMAP) \\
			Department of Physics \& Astronomy, \\
			University of California, Davis, CA 95616, USA
			\\
			\vskip 0.5cm
			$^{2}$
			Center for Theoretical Physics \\
			Massachusetts Institute of Technology\\
			Cambridge MA 02139, USA
			\\
			\vskip 0.5cm
			\tt \href{mailto:ahfirat@ucdavis.edu}{ahfirat@ucdavis.edu}, \href{mailto:raji@mit.edu}{raji@mit.edu} \vskip 5pt }
		
		\vskip 2.5cm
		{\bf Abstract}
		
	\end{center}
	\vskip 0.5cm
	\noindent
	\begin{narrower}
		\baselineskip15pt
		We supplement the string field theory action with boundary terms to make its variational principle well-posed. Central to our considerations is the violation of the stress-energy tensor conservation in non-compact CFTs due to the boundary terms. This manifests as the failure of the cyclicity of the BRST operator, which encodes the target space integration by parts identities at the level of the worldsheet. Using this failure, we argue that the free closed string field theory action admits a well-posed variational principle upon including an additional boundary contribution. We explicitly work out the resulting action up to the massless level and show that it is related to the expansion of the low-energy effective string action endowed with the Gibbons-Hawking-York term on a flat background. We also discuss the structure of the boundary terms in the interacting theory.
	\end{narrower}
\end{titlepage}

\tableofcontents

\baselineskip15pt

\section{Introduction}

String field theory (SFT) is a second-quantized formulation of string theory.\footnote{For reviews on SFT, refer to~\cite{Sen:2024nfd,Zwiebach:1992ie,Erbin:2021smf,Erler:2019loq,Maccaferri:2023vns,deLacroix:2017lif}. Although our discussion is generally applicable, we exclusively work with closed bosonic strings. We consider a 2d CFT consisting of a $c = 26$ bosonic matter CFT and the ordinary $bc$ ghost system with central charge $c = -26$, and impose level-matching on states. We take the non-compact part of the matter CFT to consist of $D$ free scalars $X^\mu$, $\mu = 0, \dots, D-1$, describing $D$ flat, non-compact directions in the target space, unless stated otherwise. We set $\alpha' = 1$.} As a second-quantized theory, it reinterprets the BRST conditions on the worldsheet
\begin{align} \label{eq:1.1}
	Q_B | \Psi \rangle = 0 
	\quad \quad \text{and} \quad \quad
	| \Psi \rangle \simeq |\Psi \rangle + Q_B | \Lambda \rangle \, ,
\end{align}
which an on-shell string mode $\Psi$ satisfies as the equation of motion and gauge symmetry on the target space at the non-interacting level. The gauge-invariant action for this theory is given by
\begin{align} \label{eq:1.2}
	S_{bulk} = {1 \over 2} \, \langle \omega | \Psi \otimes Q_B \Psi \, ,
\end{align}
where $\langle \omega |$ is the suitably defined symplectic form for which the BRST operator $Q_B$ is~\emph{cyclic}~\cite{Erler:2019loq}
\begin{align} \label{eq:1.4}
	\langle \omega | \left(
	\id \otimes Q_B + Q_B \otimes \id
	\right)
	= 0 \, .
\end{align}
Using this property, together with the graded anti-symmetry and non-degeneracy of $\langle \omega |$, it can be shown that the variation of the action~\eqref{eq:1.2} takes the form
\begin{align} \label{eq:1.5}
	\delta S_{bulk} 
	=  {1 \over 2} \langle \omega | \delta \Psi \otimes Q_B \Psi +
	{1 \over 2 } \langle \omega | \Psi \otimes Q_B \delta \Psi
	= \langle \omega | \delta \Psi \otimes Q_B \Psi \, ,
\end{align}
and from which the equation of motion~\eqref{eq:1.1} follows. Refer to section~\ref{sec:3} for further details on the construction of the SFT action. 

The procedure outlined above, however, is completely agnostic about the contributions from the non-compactness of the target space, and the variational principle for SFT is ill-posed as a result. Our primary objective in this note is to supplement the free action~\eqref{eq:1.2} with appropriate boundary terms, ensuring that the variational problem becomes well-defined. These boundary terms are expected to provide non-trivial contributions to the on-shell actions in closed SFT~\cite{Erler:2022agw} and also play a role in the stringy counterpart of the computation of black hole entropy~\cite{Gibbons:1976ue}.

More precisely, the variational principle of the action~\eqref{eq:1.2} is well-posed only for the compact part of the worldsheet CFT. To see this, we remind that the BRST operator takes the form
\begin{align} \label{eq:1.6}
	Q_B &= 
	\oint j_B \, dz + \oint \overline{j}_B \, d \overline{z}
	\\ \nonumber
	&=
	c_0^+ L^{m+}_0  - 2 c_0^+ + \cdots 
	= {1 \over 2}   c_0^+ ( p^2 - 4) + \cdots \, ,
\end{align}
for closed strings, see~\eqref{eq:2.15}. This shows that $Q_B \Psi$ contains a second derivative when expressed in position space, and the cyclicity~\eqref{eq:1.4} essentially encodes the integration by parts identities at the level of the worldsheet. This is only partially true, however, as the boundary terms that would result from integration by part are absent in~\eqref{eq:1.4}. Consequently any surface contributions arising from varying~\eqref{eq:1.5} are ignored. 

This wasn't a problem for the compact CFT portion, obviously; however these terms are crucial to obtaining a well-defined variational principle for the portion describing the non-compact target space. Since the action~\eqref{eq:1.2} contains terms of the form $\phi \partial^2 \phi$ for some of the target space fields $\phi$, thanks to~\eqref{eq:1.6}, we definitely need to supplement the action~\eqref{eq:1.2} with appropriate boundary terms to obtain an action without them, hence a variational principle. They are supposed to arise from how the BRST operator $Q_B$ fails to be cyclic under the symplectic form $\langle \omega |$.

To better probe the heart of the issue, we first revisit the derivation of cyclicity~\eqref{eq:1.4} in this paper. As we review in section~\ref{sec:3}, cyclicity is essentially a consequence of the contour deformation of the holomorphic BRST current $j_B(z)$ (and its antiholomorphic counterpart) on the sphere. This suggests there should be an obstruction to performing such contour deformations in non-compact CFTs. Recall that these contour deformations are possible as a result of the conservation law $\langle \overline{\partial} j_B \cdots \rangle = 0$, so the resolution seems to be that this must fail due to boundary contributions. 

Indeed, this is the case. In~\cite{Kraus:2002cb}, the authors argued that the certain conservation laws fail to hold in the presence of a zero mode in the worldsheet CFT. We summarize their results in section~\ref{sec:2}. Subsequently applying them to the ``holomorphic'' BRST current shows that
\begin{align} \label{eq:1.7}
	\big\langle \overline{\partial} j_B \cdots \big\rangle = - { \pi \over V}
	\int d^D x \, \partial_\mu 
	\langle D^\mu \cdots \rangle'
	\, , \quad \quad
	D^\mu \equiv c \, \partial X^\mu
	\, ,
\end{align}
where $V$ is the volume of the worldsheet and the prime on the correlator denotes the CFT correlator whose zero mode $x^\mu$ integration is stripped from the rest of the Polyakov path integral, see~\eqref{eq:2.16}. They are to be evaluated by removing $x^\mu$ integration from the integral and taking $x^\mu$ as a background field for the remaining integral over the massive modes. 

The equation~\eqref{eq:1.7} implies that the cyclic property~\eqref{eq:1.4} of the BRST operator $Q_B$ actually fails by boundary terms upon using the divergence theorem. After some manipulation, we can show
\begin{align} \label{eq:1.8}
	\langle \omega | \left( \id \otimes Q_B + Q_B \otimes \id \right) 
	= 
	{1 \over 2} \int d^D x \, \partial_\mu
	\langle \omega' | \left( \id \otimes D_B^\mu + D_B^\mu \otimes \id \right) \, ,
\end{align}
where $D_B^\mu = D_0^\mu + \overline{D}_0^\mu$ is the combination of zero modes of $D^\mu$ for closed strings and $\langle \omega' |$ denotes that the correlator function defining the symplectic form should be evaluated without $x^\mu$ integration, see~\eqref{eq:3.17}. The right-hand side of~\eqref{eq:1.8} is manifestly a boundary term.

In light of this result, we propose that the free SFT action should be supplemented by the following boundary contributions
\begin{align} \label{eq:1.8a}
	S &= 
	S_{bulk} + S_{bnd} =
	{1 \over 2} \langle \omega | \Psi \otimes Q_B \Psi 
	- {1 \over 4} \int d^D x \, \partial_\mu \langle \omega' | \Psi \otimes D_B^\mu \Psi
	\, .
\end{align}
Subjecting the string field (but~\emph{not} its derivatives) to have a vanishing variation at the boundary, we demonstrate that the variation $\delta S$ doesn't contain any boundary terms---leading to a well-posed variational principle. This, however, forces us to make a gauge choice asymptotically since the auxiliary modes in the string field $\Psi$ are required to be fixed alongside the dynamical ones. We also argue that such fixing is necessary for gauge invariance of~\eqref{eq:1.8a} on the nose. Refer to section~\ref{sec:3} for a more detailed explanation and the derivations.

In order to test the validity of our proposal for the boundary terms, we evaluate the action~\eqref{eq:1.8a} explicitly for the tachyonic and massless (graviton, Kalb-Ramond, and dilaton) fields. We show that the resulting action doesn't contain derivatives of the fields higher than the first order, which manifestly demonstrates that we have obtained the expected boundary terms. Furthermore, we compare the massless action with the expansion of the low-energy effective string action, which also contains the Gibbons-Hawking-York (GHY) surface term~\cite{York:1972sj, Gibbons:1976ue}, and confirm that they are consistent with each other upon choosing an appropriate gauge asymptotically to ensure that the variational problems are compatible.

It is important to emphasize again that having a well-posed variational principle and including the appropriate boundary terms are not just an academic exercise, but have a non-negligible impact on physics in general. This is most clearly demonstrated in general relativity. Supplementing the Einstein-Hilbert action with the GHY term~\cite{York:1972sj,Gibbons:1976ue} is necessary to obtain the correct results for the gravitational Hamiltonians~\cite{Hawking:1995fd} and black hole entropy~\cite{Gibbons:1976ue}. Since closed string theory is a gravitational theory, it is natural to anticipate that the closed SFT version of the GHY boundary term is expected to play an analogous role in the stringy counterparts of these computations. This expectation is further compounded by the recent result of Erler on the vanishing of the on-shell closed SFT action~\cite{Erler:2022agw}, based on the dilaton theorem~\cite{Bergman:1994qq}, which indicates that the only contribution to the on-shell action comes from the boundary. The boundary terms also play an important role in the covariant phase space formalism~\cite{Harlow:2019yfa, Cho:2023khj, Symp}. Finally, see~\cite{Ahmadain:2024uom,Ahmadain:2024hgd,Ahmadain:2024uyo} for a distinct approach to understand the origin of the boundary terms from the string worldsheet.

The rest of the paper is organized as follows. We begin our discussion by reviewing the arguments of~\cite{Kraus:2002cb} for the failure of the conservation laws due to boundary terms in the presence of a zero mode in the Polyakov path integral in section~\ref{sec:2}. We then use this result to show that the BRST operator fails to be cyclic under the symplectic form in section~\ref{sec:3}, which we subsequently use to argue that the action~\eqref{eq:1.8a} leads to a well-posed variational principle. In section~\ref{sec:4}, we evaluate the action~\eqref{eq:1.8a} up to the massless level and demonstrate that it doesn't contain derivatives of the fields higher than the first order. We then show that the expansion of the low-energy effective string action, endowed with the GHY term, to quadratic order in fluctuations produces the same action in section ~\ref{sec:5}. In section~\ref{sec:6}, we conclude our paper and discuss the expected features of the boundary terms in the interacting theory.

\noindent\textbf{Note added:} As we finalized this manuscript, a work on the boundary terms in the context of open string field theory has appeared in~\cite{Georg}. We have also been informed that Maccaferri et al. are working on a similar problem independently~\cite{Carlo}.

\section{The non-conservation of the BRST current} \label{sec:2}

In this section, we review the arguments of~\cite{Kraus:2002cb} for why the worldsheet stress-energy tensor (hence the BRST current) is conserved up to boundary terms. We consider the free non-compact scalars $X^\mu$, $\mu = 0, \dots, D-1$, but the arguments outlined here are generally applicable to any CFT with non-compact target spaces.

So, we begin with the Polyakov integral of the $X^\mu$ CFT in the background of $g_{ab}$
\begin{align} \label{eq:2.1}
	\langle \cdots \rangle_g =
	\int \mathcal{D} X  \, \left( e^{- I[g,X] } \cdots \right) \, ,
	\quad \quad
	I[g,X] =
	{1 \over 4 \pi} \int d^2 \sigma \sqrt{g} \, g^{ab} \eta_{\mu \nu}
	\partial_a X^\mu \partial_b X^{\nu} \, ,
\end{align}
where $\cdots$ represents the possible operator insertions. Some care must be taken to define the measure $\mathcal{D} X $ here~\cite{DHoker:1988pdl,Erbin:2021smf}. Nevertheless, since we are working with a flat background, this measure takes the form of an ordinary free field measure, which can be simply expressed as
\begin{align} \label{eq:2.2}
	\mathcal{D} X  = V^{D/2} d^D x \prod_{I \neq 0} d^D x_I 
	\equiv d^D x \,  \mathcal{D} X'
	\, ,
\end{align}
using the Fourier decomposition of $X^\mu(\sigma)$ into its harmonics $X_I(\sigma)$
\begin{align} \label{eq:2.3}
	X^\mu(\sigma) = x^\mu 
	+ \sum_{I \neq 0 } x_I^\mu X_I (\sigma) \, ,
	\quad \quad
	\nabla X_I = - \omega_I^2 X_I \, ,
\end{align}
with the normalization
\begin{align}
	\int d^2 \sigma \sqrt{g} \, X_I X_J = \delta_{ I J} \, , 
	\quad \quad I \neq 0 \, .
\end{align}
We have taken the zero mode to be $X_{0}(\sigma) = 1$ above. Note that its normalization is different from that of the massive modes and this introduces an additional worldsheet volume factor
\begin{align} \label{eq:2.5}
	V = \int d^2 \sigma \sqrt{g} \, ,
\end{align}
in the measure~\eqref{eq:2.2} as a result. We include this additional factor in the massive modes' measure. We highlight that there are no overall normalization ambiguity for the measure~\eqref{eq:2.2}: this is adjusted by renormalizing the worldsheet cosmological constant in view of total vanishing Weyl anomaly and the ultralocality of the measure, see~\cite{DHoker:1988pdl}.  

The decomposition~\eqref{eq:2.3} further allows us to decompose the functional derivative as
\begin{align}
	{\delta X^\mu (\sigma) \over \delta X^\nu (\sigma')}
	= {1 \over \sqrt{g}} \, \delta^\mu_\nu \, \delta(\sigma - \sigma') 
	\quad \implies \quad
	{\delta \over \delta X^\mu } = {1 \over V} \partial_\mu
	+ \sum_{I \neq 0 } X_I (\sigma) {\partial \over \partial x_I^\mu} \, .
\end{align}
In light of this, we find
\begin{align} \label{eq:2.7}
	\int \mathcal{D} X  \, {\delta \over \delta X^\mu } \left( e^{- I[g,X] } \cdots \right) 
	&= {1 \over V} \int \mathcal{D} X  \,  \partial_\mu \left( e^{- I[g,X] } \cdots \right) 
	= {1 \over V} \int d^D x \, \partial_\mu \langle \cdots \rangle'_g \, ,
\end{align}
given that the action $I[g,X]$ evaluates to 
\begin{align} \label{eq:2.8}
	I[g,X] = {1 \over 4 \pi} \sum_{I} {\omega_I^2 \, x_I \cdot x_{I}} \, ,
\end{align}
and the boundary terms arising from $x_I^\mu$ integration are suppressed,~\emph{except} for $I = 0$.\footnote{We assume that the rest of the insertions doe not blow up as $|x_I^\mu| \to \infty$ and the $x^\mu$ dependence only comes from $e^{i k \cdot X}$ insertions associated with the momentum.\label{fn}} The prime on the correlator indicates the integration over the zero modes $x^\mu$ is stripped, as per~\eqref{eq:2.2}, in the correlators:
\begin{align}
	\langle \cdots \rangle'_g \equiv
	\int \mathcal{D} X'  \, \left( e^{- I[g,X] } \cdots \right) \, .
\end{align}

In the standard arguments, as in~\cite{Polchinski:1998rq}, the right-hand side of~\eqref{eq:2.7} is taken to vanish (modulo contact terms). However, we see that this is not entirely true for non-compact theories---there may be non-trivial surface contributions. For our purposes, we are concerned with its effect on the conservation of the stress-energy tensor
\begin{align}
	T_{ab} = {4 \pi } \, { \delta I \over \delta g_{ab}}
	= -
	\partial_a X^\mu \partial_b X_\mu + {1 \over 2} g_{a b} \partial_c X^\mu \partial^c X_\mu
	 \, .
\end{align}
Recall that the Polyakov action $I$ is diffeomorphism-invariant so we have
\begin{align} \label{eq:2.11}
	0 = {\delta I \over \delta g_{a b}} \, \delta_\xi g_{ab} + 
	{\delta I \over \delta X^\mu} \, \delta_\xi X^\mu 
	\quad \quad \implies \quad \quad
	\nabla_a T^{a b} = 2 \pi {\delta I \over \delta X^\mu} \partial^b X^\mu \, ,
\end{align}
using the diffeomorphism
\begin{align}
	\delta_\xi g_{a b} = \nabla_a \xi_b + \nabla_b \xi_a \, , \quad \quad
	\delta_\xi X^\mu = \xi^a \partial_a X^\mu \, ,
\end{align}
and after performing the integrating by parts.

We can now read the ``conservation'' law for the stress energy tensor from~\eqref{eq:2.11}~\cite{Kraus:2002cb}
\begin{align}
	\left\langle \nabla_a T^{a b} \cdots \right\rangle_g
	= 2 \pi \left\langle {\delta I \over \delta X^\mu} \partial^b X^\mu \cdots \right\rangle_g
	&= - 2\pi \int \mathcal{D} X  \, {\delta \over \delta X^\mu (\sigma)} \left( e^{- I[g,X] } \partial^b X^\mu \cdots \right) 
	\\ \nonumber
	&= - {2 \pi \over V} \int d^D x \, \partial_\mu \left\langle \partial^b X^\mu \cdots \right\rangle'_g \, .
\end{align}
After fixing to the flat conformal gauge and switching to complex coordinates, this equality takes the form
\begin{align} \label{eq:2.14}
	\left\langle \overline{\partial} T \cdots \right\rangle
	= -{ \pi \over V }\int d^D x \, \partial_\mu \left\langle \partial X^\mu \cdots \right\rangle' \, ,
\end{align}
and similarly for the anti-holomorphic counterpart. We see that the holomorphic stress-energy tensor is no longer holomorphic and the failure is given by a boundary term. The conformal gauge dependence will eventually drop out as we shall see.

Using this result, we can also express the failure of the BRST current $j_B$ to be holomorphic. Recall that the BRST current $j_B$ is given by
\begin{align} \label{eq:2.15}
	j_B  
	 = c \, T^{X} + c \, T^{c} + b \, c \, \partial c + {3 \over 2} \partial^2 c \, ,
\end{align}
where $T^X$ and $T^c$ stand for the stress-energy tensors of the non-compact $X^\mu$ CFT and the remaining compact part of the matter CFT respectively. We use $T^m = T^X + T^c$ for the stress-energy tensor of the combined matter CFT and $T$ for the total stress-energy tensor henceforth. The failure of $\overline{\partial} j_B = 0$ entirely comes from $T^X$, so we have
\begin{align} \label{eq:2.16}
	\left\langle \overline{\partial} j_B \cdots \right\rangle
	= - {\pi \over V} \int d^D x \, \partial_\mu \left\langle D^\mu \cdots \right\rangle' \, , 
	\quad \quad D^\mu \equiv c \, \partial X^\mu
	\, .
\end{align}
This result directly follows from~\eqref{eq:2.14} and the fact that the $c$-ghost is still holomorphic. We point out that the form of the result~\eqref{eq:2.16} is background-independent in the sense that if we were using a different non-compact CFT, the only change would be in how the Grassmann-odd, weight 0 operator $D^\mu$ is defined. We don't need to consider the local coordinates for the insertions of the operator $D^\mu = c \partial X^\mu$ since it is a weight-0 primary.

Despite the fact that the conservation laws for the stress-energy tensor and BRST current are modified by boundary contributions, we emphasize that the target space translation current remains unchanged---$\partial X^\mu(z)$ is still holomorphic since
\begin{align} \label{eq:2.17}
	{1 \over \pi } \langle \overline{\partial} \partial X^\mu \cdots \rangle
	&= -  \left\langle {\delta I \over \delta X_\mu} \cdots \right\rangle
	\\ \nonumber 
	&= \int \mathcal{D} X  \, {\delta \over \delta X_\mu} \left( e^{- I[X] } \cdots \right) 
	\\ \nonumber
	&= {1 \over V} \int d^D x \, \partial_\mu \int \mathcal{D} X'  \, \left( e^{- I[X] } \cdots \right) 
	= {1 \over V} \int d^D x \, \partial_\mu \langle \cdots \rangle' \, 
	\, ,
\end{align}
modulo contact terms. The Polyakov action $I$ has no dependence on $x^\mu$ (see~\eqref{eq:2.8}) but the insertions depend on $x^\mu$ as we pointed out in footnote~\ref{fn}. Then we have
\begin{align}
	{1 \over \pi } \left\langle \overline{\partial} \partial X^\mu \mathcal{O}_1 \cdots \mathcal{O}_n \right\rangle 
	&= {i \over V} \left( \sum_{i=1}^n k_i^\mu \right) \langle \mathcal{O}_1 \cdots \mathcal{O}_n \rangle 
	\propto \left( \sum_{i=1}^n k_i^\mu \right)  \delta^{(D)} \left( \sum_{i=1}^n k_i  \right)   = 0 \, ,
\end{align}
where the operator $\mathcal{O}_i$ is taken to carry momentum $k_i$. 

We almost have the same result for the primed correlators, except that there is no delta function to impose the momentum conservation
\begin{align} \label{eq:2.19b}
	\langle \overline{\partial} \partial X^\mu \mathcal{O}_1 \cdots \mathcal{O}_n\rangle' 
	= {i \pi \over V}  \left( \sum_{i=1}^n k_i^\mu \right)  \langle \mathcal{O}_1 \cdots \mathcal{O}_n \rangle' \, ,
\end{align}
which means that $\partial X^\mu$ actually fails to be holomorphic by a total momentum factor in the primed correlators. This result will be important in section~\ref{sec:3}.

\subsection{Evaluation of the primed correlators} \label{sec:2.1}

Before we conclude this section, let us comment on the evaluation of the primed correlators within the operator formalism. For this, we first adopt the normalization
\begin{align} \label{eq:2.19a}
	\langle k' | c_{-1} \overline{c}_{-1} c_0^+ c_0^- c_1 \overline{c}_1 | k'\rangle 
	&= {1 \over 2} (2 \pi)^D \delta^{(D)} (k'+k) 
	= {1 \over 2} \int d^D x \, e^{i (k'+k) x} \, ,
\end{align}
to evaluate the ordinary correlators, where $| k \rangle$ is the vacuum carrying momentum $k$ and $c_n$'s are the modes of the $c$-ghost. Note that we wrote an integral representation of the delta function in the second equality. This suggests that the primed correlators should be evaluated using
\begin{align} \label{eq:2.18}
	\langle k' | c_{-1} \overline{c}_{-1} c_0^+ c_0^- c_1 \overline{c}_1 | k \rangle' 
	= {1 \over 2} e^{i (k'+k)x} \, ,
\end{align}
according to the separation of the zero modes $x^\mu$~\eqref{eq:2.2}. Again, we emphasize that there will be no ambiguity in this separation due to our comments below~\eqref{eq:2.5}: no ambiguity will remain in the expressions once the normalization is set according to~\eqref{eq:2.19a}.

A particularly subtle primed correlator is the following one
\begin{align} \label{eq:2.19}
	  \langle k' | c_{-1} \overline{c}_{-1} c_0^+ c_0^- c_1 \overline{c}_1 p^\mu | k \rangle'
	= {1 \over 2}k^\mu  \, e^{i (k+k')x} 
	=  - {1 \over 2}(k')^\mu  \, e^{i (k+k')x} \, ,
\end{align}
with an insertion of momentum operator $p^\mu$---the zero mode of the translation current on the worldsheet $i \partial X^\mu$. We have taken $p^\mu$ to be acting on the ket (bra) in the first (second) equality above. The relative sign is introduced as a result of the inversion. 

This naively suggests that the correlator is non-zero only when $k + k' = 0$, similar to its unprimed counterpart. However, this is not correct, as the arguments leading to the BPZ conjugation of the momentum operator have to get modified slightly in view of the result~\eqref{eq:2.19b} described earlier. In order to avoid ambiguity in expressions, we adopt the convention that the momentum operator $p^\mu$ is evaluated in the primed correlators where it appears moving forward. That is, we don't allow deforming the contour integral of $i \partial X^\mu$---given that it can pick up contributions from the background field $x^\mu$ on the worldsheet as it gets deformed due to~\eqref{eq:2.19b}. This would give rise to an ambiguity. A more detailed explanation will be given in the next section.

\section{The boundary terms in free closed SFT} \label{sec:3}

In this section, we include a suitable boundary term to the free SFT, which will make its variational problem well-posed. To do this, we first revisit the derivation of the cyclicity of the BRST operator in light of the modified ``conservation'' law~\eqref{eq:2.16} derived in the previous section.

\subsection{The failure of the cyclicity of the BRST operator}

It is beneficial to begin our discussion by reminding ourselves of the steps that go into the construction of the bulk part of the SFT action~\eqref{eq:1.2}. As we mentioned in the introduction, we are concerned with closed bosonic strings whose underlying worldsheet CFT consists of the non-compact $X^\mu$ CFT, some compact matter CFT with central charge $26-D$, and the $bc$ ghost system with central charge $c=-26$. We further impose the level-matching condition
\begin{align} \label{eq:3.1}
	| \Psi \rangle \in \widehat{\mathcal{H}} \subseteq \mathcal{H} \, ,
	\quad \quad
	L_0^- | \Psi \rangle =( L_0 - \overline{L}_0) | \Psi \rangle = 0\, , 
	\quad \quad
	b_0^- | \Psi \rangle =( b_0 - \overline{b}_0) | \Psi \rangle = 0 \, , 
\end{align}
on string fields for the reasons that will be apparent soon. Here $L_n$ ($b_n$) stands for the modes of the total stress-energy tensor $T$ ($b$-ghost).

We need a symplectic form $\langle \omega|$ to write down the bulk part of the SFT action. Before we define it, we first introduce the BPZ inner product
\begin{align} \label{eq:3.2}
	\langle \text{bpz} |: \mathcal{H} \otimes \mathcal{H} \to \mathbb{C}^{1 | 1} \, ,
	\quad \quad
	\langle \text{bpz} | \Psi_1 \otimes \Psi_2 
	= \langle \Psi_1 | \Psi_2 \rangle
	= \left\langle \Psi_1 (\widetilde{z} = 0) 
	\Psi_2 (z=0) \right\rangle \, .
\end{align}
Here $\widetilde{z}  = 1/z$ is the inversion map and we only wrote the holomorphic arguments of the operator insertions for simplicity. The form $\langle \text{bpz} |$ is non-degenerate, has even statistics, and carries an intrinsic ghost number $-6$. The following symmetry property is immediate
\begin{align} \label{eq:3.3}
	\langle \text{bpz} | \Psi_1 \otimes \Psi_2  = (-1)^{\Psi_1 \Psi_2} \langle \text{bpz} | \Psi_2 \otimes \Psi_1 \, , 
\end{align}
after using the invariance of the correlator~\eqref{eq:3.2} under inversion.

We can also establish the following exchange property
\begin{align} \label{eq:3.4}
	\langle \text{bpz} | \left(\id \otimes \phi_n - (-1)^{d} \phi_{-n} \otimes \id \right) = 0  \, ,
	\quad \quad
	\phi(z) = \sum_{n \in \mathbb{Z}} {\phi_n \over z^{n+d}} \, ,
\end{align}
for the modes of a weight $d$ primary operator $\phi(z)$ under $\langle \text{bpz} |$. We remind the reader that the derivation of~\eqref{eq:3.4} exploits the fact $\phi(z)$ is holomorphic so that the contour integrals defining the modes $\phi_n$ can be deformed freely, which eventually leads to~\eqref{eq:3.4}. As we have seen in the previous section, however, naively holomorphic operators (such as BRST current) may fail to be holomorphic due to the boundary contributions. That implies that the right-hand side of~\eqref{eq:3.4} is a boundary term in general. 

For the particular case of the BRST charge $Q_B$, we have
\begin{align} \label{eq:3.5}
	\langle \zeta | : \mathcal{H} \otimes \mathcal{H} \to \mathbb{C}^{1|1} \, ,
	\quad \quad
	\langle \text{bpz} | \left( \id \otimes Q_B + Q_B \otimes \id \right) = 
	\langle \zeta | \neq 0 \, ,
\end{align}
and the right-hand side is given by
\begin{align} \label{eq:3.6}
	\langle \zeta | \Psi_1 \otimes \Psi_2 
	&= 
	 {1 \over 2 \pi}  \int d^2 z \,  \left\langle \left( \overline{\partial} j_B (z, \overline{z})
	+ \partial \overline{j}_B (z, \overline{z}) \right)
	\Psi_1 (\widetilde{z} = 0) \, \Psi_2 (z=0) \right\rangle
	\\ \nonumber 
	&=
	- {1 \over 2V}
	\int d^D x \, \partial_\mu  \left(
	\int d^2 z \,
	\left\langle 
	\left(D^\mu(z) 
	+ \overline{D}^\mu(\overline{z}) \right)
	\Psi_1(\widetilde{z}=0) 
	\Psi_2(z=0) 
	\right\rangle'
	\right) 
	\, ,
\end{align}
upon using the complex divergence theorem
\begin{align} \label{eq:3.7}
	{1 \over 2 \pi} \int_R d^2 z \left( \partial v^z + \overline{\partial} v^{\overline{z} }\right)
	= \oint_{\partial R} \left( v^z d \overline{z} + v^{\overline{z}} d z\right) \, ,
\end{align}
to perform the contour deformation and~\eqref{eq:2.16} respectively. We adopt the conventions
\begin{align}
	\oint {dz \over z} = \oint {d \overline{z} \over \overline{z} } = 1 \, ,
\end{align}
for the contour integrals surrounding the zero. Since the original contour integrals of $j_B$ are infinitesimally close to the operator insertions, the $z$-integral is over the entire worldsheet $R$. The failure $\langle \zeta | $ has odd statistics and carries an intrinsic ghost number of $-5$. We point out that using $\partial X^\mu$ instead of $D^\mu = c \partial X^\mu$ in~\eqref{eq:3.6} similarly leads to the failure of the cyclicity of the modes of the stress-energy tensor. 

We can use the $PSL(2,\mathbb{C})$ symmetry to evaluate the integral over the worldsheet in~\eqref{eq:3.6} by setting the position of $D^\mu$ to a convenient point (say $z=1$). That is,
\begin{align} \label{eq:3.9}
	\langle \zeta | \Psi_1 \otimes \Psi_2  
	&= - {1\over 2}\int d^D x \, \partial_\mu  
	\left\langle 
	\left(D^\mu(z=1) 
	+ \overline{D}^\mu(\overline{z}=1) \right)
	\Psi_1(\widetilde{z}=0) 
	\Psi_2(z=0) 
	\right\rangle' 
	\\ \nonumber
	&= - {1 \over 2} \int d^D x \, \partial_\mu  
	\bigg\langle
	\left( \oint {D^\mu(z) \over z - 1} dz 
	+ \oint {\overline{D}^\mu(\overline{z}) \over \overline{z}-1} d\overline{z}  \right) 
	\Psi_1(\widetilde{z}=0) 
	\Psi_2(z=0) 
	\bigg\rangle' \, .
\end{align}
We don't pick up any extra factors from applying  $PSL(2,\mathbb{C})$  transformations while evaluating the integral since $D^\mu$ is a weight-zero primary. 

Note that the worldsheet volume factor has canceled above. In fact, the primed correlators should be evaluated according to~\eqref{eq:2.18}, which contains no volume factors; thus, the entire volume dependence in $\langle \zeta |$ drops out. This is reassuring: the standard formulation of SFT doesn't require endowing the worldsheet with a metric and this will continue to be the case when we include boundary terms, at least for the free theory. We also very crucially point out that adopting~\eqref{eq:2.18} to evaluate the primed correlators implies that the modes of $D^\mu$ containing momentum in~\eqref{eq:3.9} have to vanish, given that they are acting on the vacuum. 

We have expressed the $D^\mu$ insertions as contour integrals around $z=1$ in the second line of~\eqref{eq:3.9}. We now want to deform these contour integrals to wrap around $z = 0$ and $z = \infty$ instead. Like before, we need to be careful when deforming the contours in view of~\eqref{eq:2.19b}. Assuming $\Psi_i$ carries momentum $k_i$, the complex divergence theorem~\eqref{eq:3.7} produces the following extra piece
\begin{align} \label{eq:3.10}
	&-  {1\over 2} {i \pi \over V} {1 \over 2\pi} \int d^2 z \, (k_1^\mu + k_2^\mu)
	\left\langle \left( c(z) + \overline{c}(\overline{z}) \right) \Psi_1(\widetilde{z}=0) 
	\Psi_2(z=0)  \right\rangle' 
	\\ \nonumber
	&\hspace{1in}= - {i \over 4} \left( k_1^\mu + k_2^\mu \right)
	\left\langle \left( c(z=1) + \overline{c}(\overline{z}=1) \right) \Psi_1(\widetilde{z}=0) 
	\Psi_2(z=0)  \right\rangle' \, ,
\end{align}
on top of the contributions from the correlators where the contour integrals of $D^\mu$ wrap around $z=0$ and $z=\infty$. We have evaluated the worldsheet integral over the entire $z$-plane using $PSL(2, \mathbb{C})$ symmetry and the volume factors are canceled again. 

Now, analogous to~\eqref{eq:3.9}, we can write the $c$-ghost insertions at $z=1$ as contour integrals and deform them to wrap around $z=0$ and $\infty$ instead---but this time we are not going to pick up any factors from the complex divergence theorem since $c(z)$ is truly holomorphic. Upon doing so, we see that we can actually represent the piece~\eqref{eq:3.10} as coming from the operator $-i c_0^+ p^\mu/2$ acting on each state (the sign comes from the weight of the $c$-ghost). Intuitively, these terms naturally compensate for the momentum part that went missing in the correlators~\eqref{eq:3.9} due to adopting~\eqref{eq:2.18}. This was somewhat expected, see our remarks at the end of subsection~\ref{sec:2.1}.

Then we can simply write our expression as
\begin{align} \label{eq:3.11a}
	\langle \zeta | 
	= \langle \text{bpz} | \left( \id \otimes Q_B + Q_B \otimes \id \right) 
	= 
	{1 \over 2} \int d^D x \, \partial_\mu
	\langle \text{bpz}' | \left( \id \otimes D_B^\mu + D_B^\mu \otimes \id \right) \, ,
\end{align}
after the rest of the contour deformations are performed. Here we have introduced the operator 
\begin{align} \label{eq:3.11b}
	D_B^\mu =  D_0^\mu + \overline{D}_0^\mu
	= - {i \over \sqrt{2} } \left[  \sum_{n \in \mathbb{Z}} c_n \alpha_{-n}^\mu + \overline{c}_n \overline{\alpha}_{-n}^\mu \right]
	= - i c_0^+ p^\mu + \cdots \, ,
\end{align}
where $\alpha_n^\mu$ are the modes of $ i \sqrt{2} \partial X^\mu $ and the primed version of the BPZ product is defined as
\begin{align} \label{eq:3.13a}
	\langle \text{bpz}' | \Psi_1 \otimes \Psi_2 = \langle \Psi_1 (\widetilde{z} = 0)  \Psi_2 (z = 0) \rangle' \, ,
\end{align}
The form $\langle \text{bpz}' | $ has the symmetry property~\eqref{eq:3.3}. One can easily convince oneself that~\eqref{eq:3.11a} does not vanish for generic momenta, see~\eqref{eq:2.19} and the discussion thereof.

In order to define the closed SFT action, we also need to introduce the non-degenerate symplectic form. This is given by
\begin{align} \label{eq:3.14}
	\langle \omega | : \widehat{\mathcal{H}} \otimes \widehat{\mathcal{H}} \to \mathbb{C}^{1|1} \, ,
	\quad \quad
	\langle \omega |  = \langle \text{bpz} | \id\otimes c_0^-
	= - \langle \text{bpz} | c_0^-\otimes \id  \, .
\end{align}
We define the symplectic form only for the level-matched states~\eqref{eq:3.1} to make it non-degenerate. Being a symplectic form, it obeys
\begin{align} \label{eq:3.11}
	\langle \omega | \Psi_1 \otimes \Psi_2 = 
	- (-1)^{\Psi_1 \Psi_2} \langle \omega | \Psi_2 \otimes \Psi_1 \, ,
\end{align}
which directly follows from~\eqref{eq:3.3}. One can similarly introduce the primed version of the symplectic form, which has the same property.

As before, the analog of the property~\eqref{eq:3.4} for $\langle \omega |$ (i.e.,~\emph{cyclicity}) would fail for the BRST operator $Q_B$. Focusing on the failure part in particular, we have
\begin{align} \label{eq:3.12}
	\langle \omega | \id \otimes Q_B &=
	-\langle \text{bpz} |  ( c_0^- \otimes \id ) (\id \otimes  Q_B)
	\\ \nonumber
	&=\langle \text{bpz} |  (\id \otimes  Q_B) ( c_0^- \otimes \id ) 
	\\ \nonumber
	&=  \langle \zeta | c_0^- \otimes \id   - \langle \text{bpz} |  (Q_B \otimes  \id) ( c_0^- \otimes \id ) 
	\\ \nonumber
	&= \langle \zeta | c_0^- \otimes \id   -
	\langle \omega | Q_B \otimes \id \, ,
\end{align}
by~\eqref{eq:3.5}. The derivation of the last line can be found in~\cite{Erler:2019loq} and one can easily convince oneself that the arguments don't get modified. So we find
\begin{align} \label{eq:3.17}
	\langle \omega | \left( \id \otimes Q_B + Q_B \otimes \id \right)
	&= {1 \over 2}\int d^D x \, \partial_\mu
	\langle \text{bpz}' | \left( \id \otimes D_B^\mu + D_B^\mu \otimes \id \right) \left( c_0^- \otimes \id   \right)
	\\ \nonumber
	&= {1 \over 2} \int d^D x \, \partial_\mu
	\langle \omega' |  \left( \id \otimes D_B^\mu + D_B^\mu \otimes \id \right) \, , 
\end{align}
as already claimed in~\eqref{eq:1.8}. Note that we anti-commuted $c_0^-$ through $D_B^\mu$ above. 

\subsection{The action with the boundary term}

We are now ready to express the free closed SFT action with the appropriate boundary term. We propose that the action with a well-posed variational principle is given by
\begin{align} \label{eq:3.13}
	S &= S_{bulk} + S_{bnd} = {1 \over 2} \langle \omega | \Psi \otimes Q_B \Psi 
	- {1 \over 4} \int d^D x \, \partial_\mu \langle \omega' | \Psi \otimes D_B^\mu \Psi  \, ,
\end{align}
where $\Psi \in \widehat{\mathcal{H}}$ is an even string field. For the classical action we restrict this string field to have ghost number $2$.  Our claim is that $\delta S_{bnd}$ cancels against the term that is induced from the non-cyclicity of the BRST operator in $\delta S_{bulk}$, leaving a boundary term that is only proportional to $\delta \Psi $, but not its derivatives, i.e., $p^\mu \delta \Psi$.

 This can be proven as follows. We have
\begin{align} \label{eq:3.15}
	\delta S
	&=  {1 \over 2} \langle \omega | \delta \Psi \otimes Q_B \Psi
	+{1 \over 2 } \langle \omega | \Psi \otimes Q_B \delta \Psi
	\\ \nonumber
	&\hspace{0.5in} -  {1 \over 4}\int d^D x \, \partial_\mu  \langle \omega' | \delta \Psi \otimes D_B^\mu \Psi 
	-  {1 \over 4}\int d^D x \, \partial_\mu  \langle \omega' | \Psi \otimes D_B^\mu \delta \Psi \, .
\end{align}
We can rewrite the second term as
\begin{align} \label{eq:3.16}
	{1 \over 2 } \langle \omega | \Psi \otimes Q_B \delta \Psi
	&= {1 \over 4} \int d^D x \, \partial_\mu
	\langle \omega' |  \left( \Psi \otimes D_B^\mu \delta \Psi + D_B^\mu \Psi \otimes \delta \Psi \right)   
	+ {1 \over 2} \langle \omega | \delta \Psi \otimes Q_B \Psi \, ,
\end{align}
using the identities~\eqref{eq:3.17} and~\eqref{eq:3.11} respectively. The first term above cancels the last term in~\eqref{eq:3.15} and we find
\begin{align} \label{eq:3.20}
	\delta S = \langle \omega | \delta \Psi \otimes Q_B \Psi 
	- {1 \over 2} \int d^D x \, \partial_\mu  \langle \omega' | \delta \Psi \otimes  D_B^\mu \Psi 
	\, ,
\end{align}
using the graded anti-symmetry of $\langle \omega' |$. The boundary term is proportional to $\delta \Psi$ and it disappears upon imposing $\delta \Psi = 0$ at the spatial boundary, except for the contributions at the temporal boundary at $t = \pm \infty$.\footnote{This condition is actually stronger than necessary, as one may impose the weaker condition where the boundary terms become total derivatives at the spatial boundary~\cite{Harlow:2019yfa}. However we are not going to be concerned with this possibility. In fact, we set $\delta \Psi = 0$ even at the temporal boundary for reasons that will become apparent soon.\label{fn:3}} The equation of motion, $Q_B \Psi = 0$, follows from the bulk variation upon taking $\Psi$ to be constant at the boundary.

Note that we are implicitly making a gauge choice at the boundary by holding the~\emph{entire} $\Psi$ constant there. This is because $\Psi$ contains numerous auxiliary fields in addition to the dynamical modes as its components and taking all of their variations to zero at the boundary amounts to specializing the field configurations to a specific gauge. For example, setting all auxiliary fields to zero is the same as imposing the Siegel gauge~\cite{Erbin:2021smf}. Hence, the variational principle of~\eqref{eq:3.13} actually gives the solutions to $Q_B \Psi = 0$ for which $\Psi$ is asymptotically automatically constrained according to some gauge.

This requirement can be further motivated by investigating the gauge transformation of~\eqref{eq:3.13}. Taking $\delta_\Lambda \Psi = Q_B \Psi$ for the variation of the string field in~\eqref{eq:3.20}, with $\Lambda \in \widehat{\mathcal{H}}$ being a Grassmann-odd gauge parameter that has one less ghost number than $\Psi$, we see
\begin{align} \label{eq:3.21}
	\delta_\Lambda S &= \langle \omega | Q_B \Lambda \otimes Q_B \Psi
	- {1 \over 2} \int d^D x \, \partial_\mu  \langle \omega' | Q_B \Lambda \otimes D_B^\mu \Psi
	\\ \nonumber
	&=  {1 \over 2} \int d^D x \, \partial_\mu \langle \omega' | D_B^\mu Q_B \Lambda \otimes \Psi 
	\, ,
\end{align}
using the identity~\eqref{eq:3.17} applied to $Q_B \Lambda \otimes \Psi$ and $Q_B^2 =0$. So the action actually fails to be gauge-invariant by boundary terms. Even though this doesn't affect the gauge invariance of the equation of motion, the result is somewhat unsatisfactory. Given our remarks regarding the variational principle requiring asymptotic gauge-fixing above, however, we are compelled to take $Q_B \Lambda$ (\emph{and} its derivatives) to vanish at the boundary since there is no gauge symmetry at the boundary. This makes the boundary term above moot and we still consider a ``gauge-invariant'' action.

Due to these arguments, we fix the entire gauge around the boundary,~\emph{including} the temporal part in our formulation, henceforth. We crucially stress here that we are actually considering inequivalent variational problems by making different asymptotic gauge choices, given that they cannot be related to each other due to the absence of gauge transformations at the boundary as mentioned above. 

It may be somewhat unsettling then that the SFT variational principle appears not to be unique and one requires this additional data regarding the behavior at the boundary to define the theory. But still, it may be possible to introduce additional auxiliary degrees of freedom to the theory at the boundary in order to write a genuinely gauge-invariant action and relate these distinct variational principles. However, this would most likely require describing the dynamics of the boundary itself from the worldsheet, which is a priori a difficult problem. We are not going to pursue this direction in this work and will content ourselves with the asymptotic gauge-fixing described above. Nonetheless, we will see that this choice still leads to a result that is consistent with the low-energy effective string action endowed with the GHY boundary term, as discussed in section~\ref{sec:5}.

Let us point out in passing that the action~\eqref{eq:3.13} can be further supplemented by the following boundary tadpole and constant terms
\begin{align} \label{eq:3.25a}
	S_{td} = \int d^D x \, \partial_\mu \langle \omega' | \Delta^\mu \otimes \Psi + S_0 \, ,
\end{align}
without interfering with the variational principle described above. Here $\Delta^\mu \in \widehat{\mathcal{H}}$ is a string field that has one more ghost number than $\Psi$ and $S_0$ is a real constant. Even though we don't provide a first-principles way of computing $\Delta^\mu$ or $S_0$ in this work, we will see in the next section that they admit natural interpretations. Finally, we note that the action~\eqref{eq:3.15} with~\eqref{eq:3.25a} is real, given that the operator $D_B^\mu = D_0 + \overline{D}_0$ is Hermitian and assuming $\Delta^\mu$ has the same reality properties as the dynamical string field $\Psi$.\footnote{The reality condition on the string field is that the Hermitian conjugate of $\Psi$ is the minus of its BPZ conjugate~\cite{Zwiebach:1992ie}.}

\section{The tachyonic and massless actions} \label{sec:4}

In this section, we check our proposal~\eqref{eq:3.13} by evaluating the closed SFT action with the boundary term~\eqref{eq:3.13} for the tachyonic and massless fields. As we shall see, we obtain the expected boundary terms. In the next section, we will observe that the boundary terms required for massless fields are precisely those obtained from the expansion of low-energy effective string action and GHY term to second-order in fluctuations.

\subsection{The tachyonic action}

We begin by evaluating the action~\eqref{eq:3.13} for the tachyon of closed SFT, which is the lowest-lying field in the level expansion. The string field relevant here is given by
\begin{align} \label{eq:4.1}
	| \Psi \rangle = 
	2 \int {d^D k \over (2 \pi)^D } \, T(k) c_1 \overline{c}_1 | k \rangle \, ,
\end{align}
where $T(k)$ are the Fourier modes of the real tachyon field $T(x)$ in momentum space, that is
\begin{align}
	T(x) = \int {d^D k \over (2 \pi)^D} \, T(k) e^{i k x} \, , \quad \quad
	T(k)^\ast = T(-k) \, .
\end{align}
The factor of 2 in~\eqref{eq:4.1} is included to achieve the canonical normalization in the action. We assume that the excitations from the compact part of the CFT do not contribute to this level or the next for simplicity. 

We can demonstrate the following
\begin{subequations}
\begin{align}
	&Q_B | \Psi \rangle  =  
	\int {d^D k \over (2 \pi)^D } \, \left(k^2 - 4\right) T(k) c_0^+ c_1 \overline{c}_1 | k \rangle + \cdots \, ,
	\\
	&D_B^\mu | \Psi \rangle = 
	\int {d^D k \over (2 \pi)^D } \, \left( - 2 i k^\mu \right)   T(k)
	c_0^+ c_1 \overline{c}_1 | k \rangle + \cdots \, .
\end{align}
\end{subequations}
by keeping terms up to the relevant order, see~\eqref{eq:1.6} and~\eqref{eq:3.11b}. Plugging these expressions to~\eqref{eq:3.13} and evaluating the correlators as discussed in subsection~\ref{sec:2.1}, we find the action for the tachyon
\begin{align}
	S &= - {1 \over 2}  \int {d^D k' \over (2 \pi)^D } \, \int {d^D k \over (2 \pi)^D } 
	\, (2\pi)^D \delta^{(D)}(k+k') \, 
	T(k') \left( {k^2} - 4 \right) T(k)
	\\ \nonumber
	&\hspace{1in} 
	+ {1 \over 2} \int d^D x \, \partial_\mu 
	\left[ \int {d^D k' \over (2 \pi)^D } \int {d^D k \over (2 \pi)^D }  \, 
	T(k') \left(  - i k^\mu \right) T(k) e^{i (k'+k) x} 
	\right] \, .
\end{align}
This action can be expressed in position space as
\begin{align} \label{eq:4.5}
	S &= {1 \over 2}
	\int d^D x \, T \left( \partial^2 + 4 \right) T
	-
	{1 \over 2} \int d^D x \, \partial_\mu \left( T \, \partial^\mu T \right)
	= \int d^D x \left(
	- {1 \over 2} (\partial T)^2 + 2
	\right) 
	\, ,
\end{align}
upon replacing $i k \to\partial$. In the last step we have canceled the total derivative and obtained a canonically normalized action for the tachyon of mass$^2$ equals to $-4$. It has a well-posed variational principle. Notice there were no auxiliary field nor gauge symmetry at this level.

\subsection{The massless action}

Now we turn our attention to the massless fields---graviton, dilaton, and Kalb-Ramond fields. This is somewhat more convoluted since the dynamical string field contains multiple components, which are given by
\begin{align} \label{eq:4.6}
	| \Psi \rangle 
	= \int {d^D k \over (2 \pi)^D} \, \bigg[
	- {e_{\mu \nu}(k) \over 2 } \, \alpha^{\mu}_{-1} \overline{\alpha}^\nu_{-1} c_1 \overline{c}_{1}
	&+ {e(k)} \, c_1 c_{-1} + {\overline{e}(k) }\, \overline{c}_1 \overline{c}_{-1}
	\\ \nonumber
	&+ {i f_\mu(k) \over \sqrt{2} } \, c_0^+ c_1 \alpha^\mu_{-1} 
	+ { i \overline{f}_\mu(k) \over \sqrt{2}} \, c_0^+ \overline{c}_1 \overline{\alpha}^\mu_{-1}
	\bigg] | k \rangle \, .
\end{align}
We adopt the same normalization as in~\cite{Sen:2024nfd} for easy comparison, which is different from the canonical normalization. We keep the string coupling ${1 / g_s}$ in front of the string field implicit.

The bulk part of this action~\eqref{eq:3.13} has already been computed in~\cite{Sen:2024nfd}, however it is better to repeat this computation to ensure that no total derivatives are missed. We begin by writing the BRST operator
\begin{align}
	Q_B &={1 \over 2} \, c_0^+
	 p^2
	+ {1 \over \sqrt{2} } \, p \cdot \left( \alpha_{-1} c_1  + c_{-1} \alpha_1 
	+\overline{\alpha}_{-1} \overline{c}_1 + \overline{c}_{-1} \overline{\alpha}_1
	\right)
	- b_0^+ \left( c_{-1} c_1 + \overline{c}_{-1} \overline{c}_1 \right)
	+\cdots \, ,
\end{align}
by keeping more terms in thee expansion~\eqref{eq:1.6}, see~\eqref{eq:2.15}. We only kept the terms relevant to this level and observed that the terms proportional to $L_0^-$, $b_0^-$ and those contributing to the mass of the field vanish.

We can apply $Q_B$ to the components of~\eqref{eq:4.6} term-by-term and get
\begin{subequations}
\begin{align}
	Q_B \, \alpha^{\mu}_{-1} \overline{\alpha}^\nu_{-1} c_1 \overline{c}_{1} | k \rangle 
	= \left(
	{1 \over 2} k^2 \alpha^{\mu}_{-1} \overline{\alpha}^\nu_{-1} c_0^+ c_1 \overline{c}_{1} 
	+ {1 \over \sqrt{2} } \, k^\mu \overline{\alpha}^\nu_{-1} c_{-1} c_1 \overline{c}_{1}
	+ {1 \over \sqrt{2} } \, k^\nu \alpha^{\mu}_{-1} \overline{c}_{-1} c_1 \overline{c}_{1}
	\right) |k \rangle \, ,
\end{align}
and
\begin{align}
	&Q_B \, c_1 c_{-1}|k \rangle = 
	\left(
	{1 \over 2} k^2 c_0^+ c_1 c_{-1} 
	 + {1 \over \sqrt{2} } \,( k \cdot \overline{\alpha}_{-1} )\overline{c}_1 c_1 c_{-1} 
	\right) |k \rangle  \, ,
	\\
	&Q_B \, \overline{c}_1 \overline{c}_{-1}|k \rangle = 
	\left(
	{1 \over 2} k^2 c_0^+ \overline{c}_1 \overline{c}_{-1} 
	+ {1 \over \sqrt{2} } \, (k \cdot \alpha_{-1} )c_1 \overline{c}_1 \overline{c}_{-1} 
	\right) |k \rangle \, ,
\end{align}
and
\begin{align}
	&Q_B \, c_0^+ c_1 \alpha_{-1}^\mu|k \rangle = 
	\left(
	{1 \over \sqrt{2}} \, k^\mu c_{-1} c_0^+ c_1 
	+ {1 \over \sqrt{2}} \, (k \cdot \overline{\alpha}_{-1}) \overline{c}_1 c_0^+ c_1 \alpha_{-1}^\mu
	- \overline{c}_{-1} \overline{c}_1 c_1 \alpha_{-1}^\mu
	\right)|k \rangle \, ,
	\\
	&Q_B \, c_0^+ \overline{c}_1 \overline{\alpha}_{-1}^\mu|k \rangle = 
	\left(
	{1 \over \sqrt{2}} \, k^\mu \overline{c}_{-1} c_0^+ \overline{c}_1 
	+ {1 \over \sqrt{2}} \, (k \cdot \alpha_{-1}) c_1 c_0^+ \overline{c}_1 \overline{\alpha}_{-1}^\mu
	- c_{-1} c_1 \overline{c}_1 \overline{\alpha}_{-1}^\mu
	\right)|k \rangle \, .
\end{align}
\end{subequations}
Combining them into a single entity, we have
\begin{align}
	&Q_B | \Psi \rangle = \int {d^D k \over (2 \pi)^D} \, 
	\bigg[
	\left( 
	- {1 \over 4} k^2 e_{\mu \nu}(k) + {1 \over 2} i k_\nu  f_\mu(k) 
	- {1 \over 2} i k_\mu \overline{f}_\nu(k)
	\right)
	\alpha^\mu_{-1} \overline{\alpha}^\nu_{-1} c_0^+ c_1 \overline{c}_1
	\\ \nonumber
	&+  \left(
	- {k^\mu e_{\mu \nu}(k) \over 2 \sqrt{2} }
	+ {k_\nu \overline{e}(k) \over \sqrt{2} }
	- {i \overline{f}_\nu(k) \over \sqrt{2}}
	\right) \overline{\alpha}_{-1}^\nu c_{-1} c_1 \overline{c}_1
	 + \left(
	- {k^\nu e_{\mu \nu}(k) \over 2 \sqrt{2} }
	- {k_\nu e(k) \over \sqrt{2} }
	+ {i f_\mu(k) \over \sqrt{2}} 
	\right) \alpha_{-1}^\mu \overline{c}_{-1} c_1 \overline{c}_1
	\\ \nonumber 
	&+ {1 \over 2}  \left( k^2 e(k) 
	+ i k^\mu f_\mu (k)\right) c_{-1} c_0^+ c_1
	+ {1 \over 2}  \left( k^2 \overline{e}(k) 
	+  i k^\mu \overline{f}_\mu (k)\right) \overline{c}_{-1} c_0^+ \overline{c}_1
	\bigg] | k \rangle \, .
\end{align}
so that the bulk part of the action for the massless fields reads
\begin{align}
	S_{bulk} = &{1 \over 2} \int {d^D k' \over (2 \pi)^D} \int {d^D k' \over (2 \pi)^D} 
	\, (2\pi)^D \delta^{(D)} (k' + k) \bigg[
	- {1 \over 8} e^{\mu \nu}(k') \left( {1 \over 2} k^2 e_{\mu \nu}(k) - i k_\nu f_\mu(k) + i k _\mu f_\nu(k) \right)
	\nonumber \\
	&- {1 \over 4} f^\mu(k') \left( {1\over 2 } i k^\nu e_{\mu \nu}(k) 
	- i k_\mu \overline{e}(k)
	+ f_\mu(k) \right)
	- {1 \over 4} \overline{f}^\nu(k') \left(- {1\over 2 } i k^\mu e_{\mu \nu}(k) 
	- i k_\mu e(k)
	+ \overline{f}_\nu(k) \right)
	\nonumber \\
	&- {1 \over 4} e(k') \left(  k^2 \overline{e}(k) +
	ik^\mu \overline{f}_\mu(k)\right)
	- {1 \over 4} \overline{e}(k') \left(  k^2 e(k) +
	ik^\mu f_\mu(k)\right)
	\bigg] \, ,
\end{align}
using the fundamental correlator~\eqref{eq:2.19a}. It was important here~\emph{not} to use the $\delta$-function (i.e., momentum conservation) to exchange the momentum,  because doing so would have corresponded to performing integration by parts, as it ignores terms proportional to the total momentum $k'+k$ (i.e., total derivatives).

The action above can be expressed in position space as follows
\begin{align} \label{eq:4.11}
	S_{bulk} &= {1 \over 8} \int d^D x \, \bigg[
	{1 \over 4} e^{\mu \nu} \partial^2 e_{\mu \nu} 
	+ {1 \over 2} e^{\mu \nu} \partial_\nu f_\mu 
	- {1 \over 2} e^{\mu \nu} \partial_\mu \overline{f}_\nu 
	- {1 \over 2} (\partial_\nu e^{\mu \nu}) f_\mu - f^2
	+ {1 \over 2} (\partial_\mu e^{\mu \nu} )\overline{f}_\nu - \overline{f}^2
	\nonumber \\
	&\hspace{1in} + e \, \partial^2 \overline{e} 
	+ (\partial^\mu \overline{e}) f_\mu- \overline{e} \, \partial^\mu f_\mu
	+ \overline{e} \, \partial^2 e 
	+ (\partial^\mu e) \overline{f}_\mu
	-  e \, \partial^\mu \overline{f}_\mu
	\bigg] 
	\\ \nonumber
	 &= {1 \over 8} \int d^D x \, \bigg[
	{1 \over 4} e^{\mu \nu} \partial^2 e_{\mu \nu} 
	+ e \, \partial^2 \overline{e}  + \overline{e} \, \partial^2 e 
	- \left( \partial_\nu e^{\mu \nu} - 2 \partial^\mu \overline{e} \right) f_\mu
	+ \left( \partial_\mu e^{\mu \nu} + 2 \partial^\nu e \right) \overline{f}_\nu
	- f^2 -\overline{f}^2
	\bigg]  \, ,
\end{align}
We point out that we have performed integration by parts for the cross terms involving the derivatives of the auxiliary fields $f_{\mu}$ and $\overline{f}_\mu$ in the second line. This induces boundary terms proportional to $f_\mu$ and $\overline{f}_\mu$, which are constant at the boundary. So these terms are simply boundary terms on the boundary itself, hence they are irrelevant in our paradigm. One can check that the last equality is the same as (4.108) of~\cite{Sen:2024nfd} upon integrating the second term by parts twice. 

To compute the boundary part of the action~\eqref{eq:3.13}, on the other hact, we need to apply $D_B^\mu$ to $\Psi$. Like before, we can do this term-by-term using the definition~\eqref{eq:3.11b}. We have
\begin{subequations}
\begin{align}
	D_B^\mu \, \alpha^{\nu}_{-1} \overline{\alpha}^\rho_{-1} c_1 \overline{c}_{1} | k \rangle 
	= \bigg( - i k^\mu \alpha^{\nu}_{-1} \overline{\alpha}^\rho_{-1} c_0^+ c_1 \overline{c}_{1}
	- {i \over \sqrt{2}}\, \eta^{\mu \nu}   \overline{\alpha}^\rho_{-1}  c_{-1} c_1 \overline{c}_{1}
	- {i \over \sqrt{2}}\,  \eta^{\mu \rho}  \alpha^{\nu}_{-1} \overline{c}_{-1} c_1 \overline{c}_{1}
	\bigg)
	| k \rangle \, ,
\end{align} 
and
\begin{align}
	&D_B^\mu \, c_{1} c_{-1} | k \rangle =
	\bigg( - i k^\mu c_0^+ c_1 c_{-1} 
	- {i \over \sqrt{2}} \, \overline{\alpha}_{-1}^\mu \overline{c}_1 c_1 c_{-1}
	\bigg)|k \rangle \, ,
	\\
	&D_B^\mu \, \overline{c}_{1} \overline{c}_{-1} | k \rangle =
	\bigg( - i k^\mu c_0^+ \overline{c}_1 \overline{c}_{-1} 
	- {i \over \sqrt{2}} \, \alpha_{-1}^\mu c_1 \overline{c}_1 \overline{c}_{-1}
	\bigg)|k \rangle \, ,
\end{align}
and
\begin{align}
	&D_B^\mu \, c_0^+ c_1 \alpha_{-1}^\nu | k \rangle 
	= \bigg(
	- {i \over \sqrt{2}}\, \eta^{\mu \nu} c_{-1} c_0^+ c_1 | k \rangle 
	- {i \over \sqrt{2}}\, \overline{c}_{1} c_0^+ c_1 \alpha_{-1}^\nu  \overline{\alpha}^\mu_{-1}
	\bigg) | k \rangle \, ,
	\\
	&D_B^\mu \, c_0^+ \overline{c}_1 \overline{\alpha}_{-1}^\nu | k \rangle =
	\bigg(
	- {i \over \sqrt{2}} \, \eta^{\mu \nu} \overline{c}_{-1} c_0^+ \overline{c}_1 | k \rangle 
	- {i \over \sqrt{2}} \, c_{1} c_0^+ \overline{c}_1 \alpha_{-1}^\mu  \overline{\alpha}^\nu_{-1}
	\bigg) | k \rangle \, .
\end{align}
\end{subequations}
Combining them into a single entity, we find
\begin{align}
	D_B^\mu | \Psi \rangle &=
	\int {d^D k \over (2 \pi)^D} \, \bigg[
	{1 \over 2} \left(i k^\mu  e_{\nu \rho}(k)  + \delta^\mu_\rho f_\nu(k)
	- \delta^\mu_\nu \overline{f}_\rho(k)
	\right) \alpha^{\nu}_{-1} \overline{\alpha}^\rho_{-1} c_0^+ c_1 \overline{c}_{1}
	\\ \nonumber
	&+ {i \over \sqrt{2}} \left( {1 \over 2}
	\eta^{\mu \nu} e_{\nu \rho}(k) + \delta^{\mu}_\rho e(k)
	\right) \overline{\alpha}^\rho_{-1}  c_{-1} c_1 \overline{c}_{1}
	+ {i \over \sqrt{2}} \left( {1 \over 2}
	\eta^{\mu \rho} e_{\nu \rho} (k)- \delta^{\mu}_\nu \overline{e}(k)
	\right)   \alpha^{\nu}_{-1} \overline{c}_{-1} c_1 \overline{c}_{1}
	\\ \nonumber
	&+ \left( - i k^\mu e(k) + {1 \over 2}  f^{\mu}(k) \right)c_{-1} c_0^+ c_1
	+ \left( - i k^\mu \overline{e}(k) + {1 \over 2} \overline{f}^{\mu}(k) \right)
	\overline{c}_{-1} c_0^+ \overline{c}_1 
	\bigg] | k \rangle \, .
\end{align}
The action $S_{bnd}$ for the massless fields then reads
\begin{align}
	S_{bnd} &= -{1 \over 4} 
	\int {d^D k' \over (2 \pi)^D} \, \int {d^D k \over (2 \pi)^D} \,
	\int d^D x \, \partial_\mu \bigg[
	{1 \over 8} \left( e^{\nu \rho} (k')   (i k^\mu) e_{\nu \rho} (k)
	+ e^{\nu \mu} (k')  f_\nu(k) -  e^{\mu \nu} (k')\overline{f}_\nu(k)
	 \right)
	 \nonumber \\
	 &\hspace{0.5in}+ {1 \over 2} e(k') \left( ik^\mu \overline{e}(k) - {1 \over 2} \overline{f}^\mu(k) \right)
	 + {1 \over 2} \overline{e}(k') \left( ik^\mu e(k) - {1 \over 2} f^\mu(k) \right)
	  \\
	 &\hspace{0.5in}- {1 \over 8} f^\nu (k') \, \eta^{\mu \rho} e_{\nu \rho}(k)
	 + {1 \over 4} f^\mu (k') \, \overline{e}(k)
	 + {1 \over 8} \overline{f}^\nu (k') \, \eta^{\mu \rho} e_{\rho \nu}(k)
	 + {1 \over 4} \overline{f}^\mu (k') \, e(k)
	\bigg] e^{i(k'+k)x} \nonumber \\
	& = -{1 \over 4} 
	\int {d^D k' \over (2 \pi)^D} \, \int {d^D k \over (2 \pi)^D} \, \int d^D x \, \partial_\mu
	\bigg[
	{1 \over 8}  e^{\nu \rho} (k')   (i k^\mu) e_{\nu \rho} (k) 
	\nonumber \\
	 &\hspace{3in}+
	{1 \over 2} e(k') (i k^\mu) \overline{e}(k)
	+ {1 \over 2} \overline{e}(k') (i k^\mu) e(k)
	\bigg] e^{i(k'+k)x} \nonumber\, ,
\end{align}
where we used the symmetry between $k$ and $k'$ in the integrals to eliminate the terms without the momentum factors. Now it is straightforward to express the boundary terms in position space as
\begin{align} \label{eq:4.15}
	S_{bnd} = {1 \over 8} \int d^D x \, \partial_\mu
	\bigg[
	- {1 \over 4} e^{\nu \rho} \, \partial^\mu e_{\nu \rho}
	- e \, \partial^\mu  \overline{e} 
	-\overline{e} \, \partial^\mu e
	\bigg] \, .
\end{align}
There are no boundary terms containing the auxiliary fields $f$ and $\overline{f}$---their contributions cancel among themselves. This was not immediately obvious from the form of the action~\eqref{eq:3.13}. 

Combining~\eqref{eq:4.11} and~\eqref{eq:4.15}, we find
\begin{align} \label{eq:4.18}
	S_{bulk} + S_{bnd}  = {1 \over 8} \int d^D x \, \bigg[
	-{1 \over 4} (\partial_\rho e_{\mu \nu})^2
	&-2 \, \partial^\mu e \, \partial_\mu \overline{e}
	\\ \nonumber
	&- \left( \partial_\nu e^{\mu \nu} - 2 \partial^\mu \overline{e} \right) f_\mu
	+ \left( \partial_\mu e^{\mu \nu} + 2 \partial^\nu e \right) \overline{f}_\nu
	- f^2 -\overline{f}^2
	\bigg]  \, ,
\end{align}
which contains no second-order derivatives acting on the fields.

We can now integrate out the auxiliary fields $f_\mu$ and $\overline{f}_\mu$ using their equations of motion
\begin{align} \label{eq:4.17}
	f_\mu + {1 \over 2} \partial^{\nu} e_{\mu \nu} - \partial_\mu \overline{e} = 0 
	\, , \quad \quad
	\overline{f}_\nu - {1 \over 2} \partial^\mu e_{\mu \nu} - \partial_\nu e = 0 \, ,
\end{align} 
from the action above and obtain
\begin{align} \label{eq:4.18a}
	S = {1 \over 8} \int d^D x \, \bigg[
	-{1 \over 4} (\partial_\rho e_{\mu \nu})^2
	&+ {1 \over 4} (\partial^\nu e_{\mu \nu})^2
	+  {1 \over 4} (\partial^\mu e_{\mu \nu})^2
	\\ \nonumber
	&
	- \partial_\nu e^{\mu \nu} \partial_\mu \overline{e}
	+ \partial_\mu e^{\mu \nu} \partial_\nu e
	+ (\partial \overline{e})^2 + (\partial e)^2
	-2 \, \partial^\mu e \, \partial_\mu \overline{e}
	\bigg] \, .
\end{align}
We highlight that the inequivalent variational principles of~\eqref{eq:3.20} become equivalent at the massless level upon integrating $f_\mu$ and $\overline{f}_\mu$ out. The action~\eqref{eq:4.18a} no longer requires fixing the values of the auxiliary fields at the boundary and they all describe the same variational principle as a result.\footnote{There is still a asymptotic gauge choice, however. This is done by keeping all the components of $e_{\mu\nu}, e ,\overline{e}$ fixed.}

Furthermore, we can set $e = - \overline{e} = d$ by making a gauge choice as discussed in~\cite{Sen:2024nfd}, and get
\begin{align}
	S = {1 \over 8} \int d^D x \, \bigg[
-{1 \over 4} (\partial_\rho e_{\mu \nu})^2
+ {1 \over 4} (\partial^\nu e_{\mu \nu})^2
&+  {1 \over 4} (\partial^\mu e_{\mu \nu})^2
\\ \nonumber
&+ \partial_\nu e^{\mu \nu} \partial_\mu d
+ \partial_\mu e^{\mu \nu} \partial_\nu d
+4 (\partial d)^2
\bigg]  \, .
\end{align}
Note that this gauge choice is consistent with holding both $e$ and $\overline{e}$ fixed, so it is permitted by the variational principle, and can be imposed everywhere---at the boundary in particular.

Finally, upon decomposing the fields into the symmetric and anti-symmetric components
\begin{align}
	&e_{\mu \nu} = h_{\mu \nu } + b_{\mu \nu} \, , \quad \quad
	h_{\mu \nu} = h_{\nu \mu} \, , \quad \quad
	b_{\mu \nu} = - b_{\nu \mu} \, , \quad \quad 
\end{align}
we can express the action above as
\begin{align} \label{eq:4.23}
	S = {1 \over 8} \int d^D x \, \bigg[
	-{1 \over 4} (\partial_\rho h_{\mu \nu})^2
	+ {1 \over 2} (\partial^\nu h_{\mu \nu})^2
	+ 2 \partial_\mu h^{\mu \nu} \partial_\nu d 
	&+ 4 (\partial d)^2
	\\ \nonumber
	&-{1 \over 4} (\partial_\rho b_{\mu \nu})^2
	+ {1 \over 2} (\partial^\nu b_{\mu \nu})^2
	\bigg] \, .
\end{align}
This form shows that the anti-symmetric part $b_{\mu \nu}$ is decoupled from the rest. We have obtained an action that only contains terms of the form $(\partial \phi)^2$ for the massless string modes. It clearly provides a well-posed variational principle.

\section{The relation to the GHY boundary term} \label{sec:5}

The final result~\eqref{eq:4.23} from the previous section can be obtained from the low-energy effective action of bosonic string theory, if we also include the Gibbon-Hawking-York (GHY) term. In this section, we demonstrate this by expanding it to second order in fluctuations on a flat, fluxless background with a boundary of  zero extrinsic curvature.

We begin by reminding that the low-energy effective string action, describing the metric $G_{\mu \nu}$, the Kalb-Ramond field $B_{\mu \nu}$, and the dilaton $\Phi$, together with the GHY boundary term, is given by
\begin{align} \label{eq:4.24}
	\mathbf{S} &= {1 \over 2 \kappa^2} \int_M d^D x \, \sqrt{-G} e^{-2 \Phi}
	\left[
	R - {1 \over 12} H^2 + 4 (\nabla \Phi)^2
	\right] 
	+ {1 \over \kappa^2 } \int_{\partial M} d^{D-1} x \sqrt{|\gamma|} e^{-2 \Phi} K 
	\, , 
\end{align}
at the leading order in the $\alpha'$ expansion. Here $\kappa$ is the gravitational constant (related to the string coupling by $\kappa = 2g_s$ in the conventions of~\cite{Sen:2024nfd}), $R$ is the Ricci scalar associated with the string frame metric $G_{\mu \nu}$, $H$ is the field strength for the two-form field $B$, $\gamma_{\mu \nu}$ is the induced metric, and $K$ is the extrinsic curvature of the boundary. It is also possible to include the counterterm to this action, but we keep it implicit for now. We consider the theory on a manifold $M$ with the boundary $\partial M$. The covariant derivative associated with $G_{\mu \nu}$ is denoted by $\nabla_\mu$.

Upon varying $\mathbf{S} $, one finds~\cite{Kraus:2002cb,Polchinski:1998rq}
\begin{align} \label{eq:5.2}
\delta \mathbf{S} = -{1 \over 2 \kappa^2} \int_M &d^D x \, \sqrt{ -G } e^{-2 \Phi} \bigg[
	\delta G_{\mu \nu} \left(\beta^{G, \mu \nu} - {1 \over 2} G^{\mu \nu} \beta^\Phi \right) 
	+  \delta B_{\mu \nu} \beta^{B, \mu \nu}
	+ 2 \delta \Phi \beta^\Phi
	\bigg]
	\\ \nonumber
	&- {1 \over 2 \kappa^2} \int_{\partial M} d^{D-1} x \, \sqrt{|\gamma|} e^{- 2 \Phi}
	\bigg[
	\delta G_{\mu \nu} \widehat{\beta}^{G, \mu \nu}
	+ \delta B_{\mu \nu} \widehat{\beta}^{B, \mu \nu}
	+ 4 \delta \Phi \widehat{\beta}^\Phi
`	- 2 \mathcal{D}_\mu c^\mu
	\bigg] \, ,
\end{align}
where we have the bulk equations of motions 
\begin{subequations} \label{eq:5.3}
\begin{align}
	&\beta^{G}_{\mu \nu}  = R_{\mu \nu} - {1 \over 2} H_{\rho \sigma \mu} H^{\rho \sigma}_{\; \; \; \; \nu} + 2 \nabla_\mu \nabla_\nu \Phi \, ,
	\\
	&\beta^{B}_{\mu \nu} = -{1\over 2} \nabla^{\rho} H_{\mu \nu \rho}
	+ \nabla^\rho \Phi H_{\rho \mu \nu} \, ,
	\\
	&\beta^{\Phi} = R - {1 \over 12} H^2 + 4 \nabla^2 \Phi - 4 (\nabla \Phi)^2 \, ,
\end{align}
\end{subequations}
and their boundary counterparts
\begin{subequations} \label{eq:5.4}
\begin{align}
	\widehat{\beta}^G_{\mu \nu} &= K_{\mu \nu} - \gamma_{\mu \nu} K + 2 \gamma_{\mu \nu} n^\rho \nabla_{\rho} \Phi \,  ,
	\\
	\widehat{\beta}^B_{\mu \nu} &= {1 \over 2} n^{\rho} H_{\mu \nu \rho} \, , \\
	\widehat{\beta}^\Phi &= K - 2 n^\mu \nabla_{\mu} \Phi \, ,
\end{align}
\end{subequations}
where $n_\mu$ is the spacelike unit vector normal to the boundary $\partial M$, directed outwards.\footnote{We can consider the spacelike portion of the boundary $\partial M$ in a similar way by considering the inward-pointing timelike unit normal vector $n^\mu$.} We have also introduced the boundary covariant derivative $\mathcal{D}_\mu$ that is compatible with $\gamma_{\mu \nu}$ above and defined
\begin{align} \label{eq:5.5}
	c^\mu = -{1 \over 2} \gamma^{\mu \nu} n^{\rho} \delta G_{\nu \rho} \, .
\end{align}
The action~\eqref{eq:4.24} provides a well-posed variational problem for the string background at the first order in the $\alpha'$ expansion upon fixing the variations of the~\emph{induced} metric $\gamma_{\mu \nu}$, longitudinal part of $B_{\mu \nu}$, and $ \Phi$ to zero (i.e., imposing the Dirichlet boundary conditions) at the spatial boundary.

In order to compare~\eqref{eq:4.24} to the action~\eqref{eq:4.23}, we need to expand~\eqref{eq:4.24} around the flat background. For this, we take
\begin{align} \label{eq:5.6}
	G_{\mu \nu} = \eta_{\mu \nu} + h_{\mu \nu} \, , \quad \quad
	B_{\mu \nu} = b_{\mu \nu} \, , \quad \quad
	\Phi = d + {1 \over 4} \eta^{\mu \nu} h_{\mu \nu} 
	= d + {1 \over 4} h\, ,
\end{align}
where we named the fluctuations suggestively. We assume that we expand the action around the ``on-shell'' background, which means that not only~\eqref{eq:5.3} are satisfied, but we also take the boundary to satisfy its ``equations of motion''~\eqref{eq:5.4}. The particular situation we are interested in is where $H$ and $\Phi$ are both turned off at the boundary, which requires $K_{\mu \nu} = 0$ by~\eqref{eq:5.4}. This background can be easily engineered considering a flat boundary with a constant normal vector $n^\mu$; however we don't need to make this choice. We discuss what happens when this condition is violated below.

It is beneficial to remark on the behavior of fluctuations at the boundary before we begin our analysis. As we argued in section~\ref{sec:3}, the variational principle of SFT requires choosing a certain gauge asymptotically, so we need to work in the appropriate gauge if we would like to obtain the equivalent principle to~\eqref{eq:4.24}. To do this, we highlight that the SFT variational principle requires all components of $h_{\mu \nu}, b_{\mu \nu}$ to be fixed, while the variational problem of~\eqref{eq:4.24} only demands fixing the pullback of the longitudinal components of $G_{\mu \nu}, B_{\mu \nu}$ (assuming the Dirichlet boundary conditions), which in turn makes the longitudinal modes of $h_{\mu \nu}, b_{\mu \nu}$ zero by~\eqref{eq:5.6} at the spatial boundary. We also demand them to be zero at the temporal boundary, since we don't want to modify the background in time through fluctuations.

So we additionally need to impose
\begin{align} \label{eq:5.7}
	n^\mu h_{\mu \nu} = n^\mu b_{\mu \nu} = 0 \, ,
\end{align}
for~\eqref{eq:5.6} at the boundary to relate the fluctuations consistently. This is an asymptotic gauge choice for $h_{\mu \nu}, b_{\mu \nu}$: we are working in the Riemann normal coordinates around the boundary with an axial gauge for the $B$-field. The derivatives of the fluctuations are not necessarily constrained.

Now, we would like to expand~\eqref{eq:4.24} to second order in fluctuations. Observe that~\eqref{eq:5.2} already provides the expansion to first order in fluctuations (which vanishes on our on-shell background), so we just need to vary~\eqref{eq:4.24} once more. First, we have
\begin{subequations} \label{eq:5.8}
\begin{align}
	\delta \beta_{\mu \nu}^G &= 
	\delta R_{\mu \nu} + 2 \partial_\mu \partial_\nu \left(
	d + {1 \over 4} h
	\right)
	= {1 \over 2}\partial^\rho \partial_\mu h_{\rho \nu}
	 - {1 \over 2} \partial^2 h_{\mu \nu} 
	+ 2 \partial_\mu \partial_\nu d \, ,
	\\
	\delta \beta_{\mu \nu}^B &= - {1 \over 2} \partial^\rho \delta H_{\mu \nu \rho}
	= - \partial^\rho \partial_\mu b_{\nu \rho} 
	-{1 \over 2} \partial^2 b_{\mu \nu} \, ,
	\\
	\delta \beta^\phi &= \delta R + 4 \partial^2 \left(
	d + {1 \over 4} h
	\right)
	= \partial^\mu \partial^\nu h_{\mu \nu} + 4 \partial^2 d 
	\, ,
\end{align}
\end{subequations}
using the variations of $R_{\mu \nu}$ and $R$ on the flat fluxless background~\cite{Wald:1984rg}. We use the expansion~\eqref{eq:5.6} for our presentation, in particular $\delta g_{\mu \nu} = h_{\mu \nu}$. Upon doing so, however, we need to be careful about the signs that may be induced from raising and lowering the indices in these expressions since $\delta g^{\mu \nu} = - h^{\mu \nu}$. We also kept the symmetries of the indices of $\beta$ implicit since we will eventually contract them with the tensors that have corresponding symmetries.

We also need
\begin{subequations}
\begin{align}
	\delta \widehat{\beta}^G_{\mu \nu} &= \delta K_{\mu \nu} 
	- \gamma_{\mu \nu} \delta K + 2 \gamma_{\mu \nu} n^\rho \partial_\rho \left(
	d + {1 \over 4} h
	\right)  \\ \nonumber
	&= - \gamma_{\mu}^{\; \; \rho }\gamma_{\nu}^{\; \; \sigma} n^\lambda \partial_\rho h_{\sigma \lambda}
	+ {1 \over 2}  \gamma_{\mu}^{\; \; \rho} \gamma_{\nu}^{\; \; \sigma} n^\lambda  \partial_\lambda h_{\rho \sigma}
	+{1 \over 2} \gamma_{\mu \nu} n^\rho \partial^\sigma h_{\rho \sigma}
	+ 2 \gamma_{\mu \nu} n^\rho \partial_\rho d 
	\\ \nonumber
	&= - n^\lambda \partial_\mu h_{\nu \lambda}
	+ {1 \over 2} n^\lambda \partial_\lambda h_{\mu \nu}
	+ {1 \over 2} \eta_{\mu \nu} n^\rho \partial^\sigma h_{\rho \sigma}
	+ 2 \eta_{\mu \nu} n^\rho \partial_\rho d \, ,
\end{align}
and
\begin{align} 
	\delta \widehat{\beta}^B_{\mu \nu}&= {1\over 2} n^\rho \delta H_{\mu \nu \rho}
	= n^\rho \partial_\mu b_{\nu \rho}  + {1 \over 2} n^\rho \partial_\rho b_{\mu \nu} \, ,
	\\
	\delta \widehat{\beta}^\Phi &= \delta K - 2 n^\mu \partial_\mu \left(
	d + {1 \over 4} h
	\right)
	=-{1 \over 2} n^\mu \partial^\nu h_{\mu \nu}
	- 2  n^\mu \partial_\mu d 
	\, ,
\end{align}
\end{subequations}
where we used $K_{\mu \nu} = 0$ for the fluxless boundary and noted $c^\mu$~\eqref{eq:5.5} is identically zero on the boundary thanks to~\eqref{eq:5.7}. The last equality of $\delta \widehat{\beta}^G_{\mu \nu}$ holds only when the $\mu,\nu$ indices are directed tangent to the boundary $\partial M$ given that $n^\mu \delta \widehat{\beta}^G_{\mu \nu} = 0$. Using this equality rather than one containing $\gamma$ will be simpler below and there won't be any difference in the final result since we will always contract it with $h^{\mu \nu}$ which satisfies $n_\mu h^{\mu \nu} = 0$, see~\eqref{eq:5.7}.

Combining all of these together, we can show that the expansion of $\mathbf{S}$ to second order in fluctuations is given by
\begin{align}
	\mathbf{S} =- {1 \over  4 \kappa^2}
	\int_M d^D x \, \bigg[
	&h_{\mu \nu} \left(
	\partial_\rho \partial^\mu h^{\rho \nu} - {1 \over 2} \partial^2 h^{\mu \nu} 
	+ 2 \partial^\mu \partial^\nu d 
	\right)
	- {1 \over 2} h \left(
	\partial^\mu \partial^\nu h_{\mu \nu} + 4 \partial^2 d 
	\right)
	\\
	&+ b_{\mu \nu} \left( 
	-\partial_\rho \partial^\mu b^{\nu \rho}
	- {1 \over 2} \partial^2 b^{\mu \nu}
	\right)
	+ 2 \left( d + {h \over 4} \right) \left(
	\partial^\mu \partial^\nu h_{\mu \nu} + 4 \partial^2 d \right)
	\bigg]
	\nonumber \\
	&\hspace{-1in} - {1 \over  4 \kappa^2}\int_{\partial M} d^{D-1} x \, \bigg[
	h_{\mu \nu} \left(
	 -n_\lambda \partial^\mu h^{\nu \lambda}
	+{1 \over 2} n^\lambda \partial_\lambda h^{\mu \nu}
	+ {1 \over 2} \eta^{\mu \nu} n^\rho \partial^\sigma h_{\rho \sigma}
	+ 2 \eta^{\mu \nu} n^\rho \partial_\rho d
	\right)
	\nonumber \\ \nonumber
	&+ b_{\mu \nu} \left(
	n_\rho \partial^\mu b^{\nu \rho}  + {1 \over 2} n_\rho \partial^\rho b^{\mu \nu}
	\right)
	+ 4 \left( d + {h \over 4} \right) 
	\left(-{1 \over 2} n^\mu \partial^\nu h_{\mu \nu}
	- 2  n^\mu \partial_\mu d \right)
	\bigg] 	+ \cdots \, ,
\end{align}
after considering the variation of~\eqref{eq:5.2} on the fluxless on-shell background and being mindful of the signs conventions mentioned below~\eqref{eq:5.8}. The additional division by $2$ comes from taking the second functional derivative.

Combining the bulk and boundary terms, we see
\begin{align} \label{eq:5.12}
	\mathbf{S} = {1 \over  2 \kappa^2}
	\int_M d^D x \, \bigg[
	-{1 \over 4} (\partial_\rho h_{\mu \nu})^2
	+ {1 \over 2} (\partial^\nu h_{\mu \nu})^2
	&+ 2 \partial_\mu h^{\mu \nu} \partial_\nu d + 4 (\partial d)^2
	\\ \nonumber
	&-{1 \over 4} (\partial_\rho b_{\mu \nu})^2
	+ {1 \over 2} \partial^\rho b^{\mu \nu} \partial_\mu b_{\rho \nu}
	\bigg] + \cdots \, .
\end{align}
Here we have also used~\eqref{eq:5.7} to eliminate a boundary term. This readily demonstrates that $(h_{\mu \nu}, d)$ part is the same as~\eqref{eq:4.23} on the nose (recall that $\kappa = 2 g_s$ and $1/g_s^2$ in front of the action~\eqref{eq:4.23} was implicit).

For the $b$-field portion, on the other hand, we have the following difference
\begin{align}
	\partial^\rho b^{\mu \nu} \partial_\mu b_{\rho \nu}
	&= (\partial^\mu b_{\mu \nu})^2   + \partial^\rho (b^{\mu \nu} \partial_\mu b_{\rho \nu})
	- \partial_\mu (b^{\mu \nu} \partial^\rho  b_{\rho \nu})
	\\ \nonumber
	&= (\partial^\mu b_{\mu \nu})^2   + \partial^\mu \left(
	b^{\rho \nu} \partial_\rho b_{\mu \nu}
	- b_{\mu \nu} \partial_\rho b^{\rho \nu}
	\right)
	\\ \nonumber
	&=  (\partial^\mu b_{\mu \nu})^2   + \partial^\mu \left(
	-2 b_{\mu \nu} \partial_\rho b^{\rho \nu} + \partial_\rho (b^{\rho \nu} b_{\mu \nu})
	\right) \, ,
\end{align} 
for its second term. At this point, we should be reminded of the condition~\eqref{eq:5.7}, which makes the second term moot, and the fact that the third term is a boundary term on the boundary, so it is irrelevant in $\mathbf{S}$. This means that we can replace $\partial^\rho b^{\mu \nu} \partial_\mu b_{\rho \nu} \to  (\partial^\mu b_{\mu \nu})^2  $~\eqref{eq:5.12}. Again, we land on the action in~\eqref{eq:4.23}.

A few remarks are in order here. We emphasize again that the SFT action with the boundary term~\eqref{eq:3.13} evaluated at the massless level~\eqref{eq:4.23} is shown to be the same as the effective string action with the GHY term~\eqref{eq:4.24} expanded to the second order in fluctuations upon imposing the gauge~\eqref{eq:5.7} asymptotically on the flat, fluxless background. A quick inspection reveals that there is not a single term that directly corresponds to the GHY boundary term from~\eqref{eq:3.13}. Instead, the entire combination of~\eqref{eq:3.13} is related to the expansion of the entirety of~\eqref{eq:4.24}.

We also highlight that we have considered $\widehat{\beta} = 0$ condition as part of the background being on-shell, but we could have just as easily imagined the scenario where $\widehat{\beta} \neq 0$. In such cases, expanding in fluctuations generally leads to boundary tadpole terms, as can be seen from~\eqref{eq:5.2}. This suggests that such boundary terms are part of $S_{tp}$ of~\eqref{eq:3.25a} with the string field $\Delta^\mu$ encodes the structure of a particular boundary from worldsheet perspective. This relationship is analogous to the relationship between the bulk tadpole in the SFT action and the off-shell string background, see the discussion in~\cite{Zwiebach:1992ie}. In a similar vein, one may attempt to connect the constant term in~\eqref{eq:3.25a} to the non-dynamical term in~\eqref{eq:4.24}. It would be desirable to derive these terms from the first principles. However, it remains unclear how to do so in this context. The ideas for evaluating the sphere amplitudes with fewer than three punctures may be helpful here, see the works~\cite{Mahajan:2021nsd,Halder:2023adw,Ahmadain:2022tew,Ahmadain:2022eso,Ahmadain:2024hdp,Erbin:2019uiz}.

\section{Discussion} \label{sec:6}

In this work, we have supplemented the free SFT action with boundary terms to endow it with a well-posed variational principle, see~\eqref{eq:3.13}. Encoding the target space integration by parts identities at the worldsheet level through the failure of the cyclicity of the BRST operator~\eqref{eq:3.17} was central to our discussion. We have explicitly investigated the resulting action explicitly at both the tachyonic level~\eqref{eq:4.5} and the massless level~\eqref{eq:4.23}, with the latter is shown to include the GHY boundary term~\eqref{eq:4.24}. Beyond the massless level, the action~\eqref{eq:3.13} also contains the appropriate boundary terms for the massive target space fields.

Beyond investigating the other types of free SFTs and various supersymmetric extensions (see \cite{Chakrabarti:2022jcb} for example), the most important next step for future work is generalizing our construction to the interacting SFT. However, the precise details of how this procedure would work are not entirely clear, although one may anticipate its general structure in light of the variational principle.\footnote{One may naturally question the applicability of the variational principle here given the non-local nature of the theory. In fact, there is some indication that the non-locality in spatial directions is irreducible~\cite{Erler:2004hv,Erbin:2021hkf}. We don't comment further on this issue in this work.} For instance, we may naturally postulate that the string products $L_{g,n}$ describing the elementary interactions would no longer be cyclic under $\langle \omega | $ now and we would instead have\footnote{Refer to~\cite{Erler:2019loq} for the construction of the interacting theory and conventions.}
\begin{align} \label{eq:6.1}
	\langle \omega | \left( \id \otimes L_{g,n} + L_{g,n} \otimes \id \right)
	= \int d^D x \, \partial_{\mu} \langle \omega' | \left( \id \otimes B_{g,n}^\mu + B_{g,n}^\mu \otimes \id \right) \, ,
\end{align}
for $n\geq1, g\geq 0$. Here the graded-symmetric maps 
\begin{align}
	B_{g,n}^\mu: \widehat{\mathcal{H}}^{\otimes n} \to \widehat{\mathcal{H}}  \, ,
\end{align}
are supposed to encode the integration by parts identities for the elementary interactions. The proposed identity~\eqref{eq:6.1} is essentially a generalization of~\eqref{eq:3.17}---we have the operator $L_{0,1} = Q_B$ and $B_{0,1}^\mu = D_B^\mu/2$.

Assuming such an identity and following similar steps in our derivation of~\eqref{eq:3.13}, we are led to consider the following interacting action
\begin{align} \label{eq:6.2}
	S = {1 \over 2!} \langle \omega | \Psi \otimes Q_B  \Psi 
	&+ {1 \over 3!} \langle \omega  | \Psi  \otimes L_{0,2}(\Psi^2) + \cdots 
	\\ \nonumber
	&- \int d^D x \, \partial_{\mu} \bigg[ 
	 {1 \over 2!}  \langle \omega' | \Psi \otimes B_{0,1}^\mu \Psi 
	 + {1 \over 3! } \langle \omega' | \Psi \otimes B_{0,2}^\mu (\Psi^2 )+ \cdots 
	 \bigg]  \, ,
\end{align}
that contains higher boundary terms so that the variations take the form
\begin{align}
	\delta S= \langle \omega | \delta \Psi \otimes \bigg( Q_B \Psi &+ {1 \over 2!} L_{0,2}(\Psi^2) + \cdots \bigg)
	\\ \nonumber
	&- \int d^D x \, \partial_{\mu} \langle \omega' | \bigg[ 
	\delta \Psi \otimes \bigg( B_{0,1} \Psi + {1 \over 2!} B_{0,2} (\Psi^2) + \cdots \bigg)
	\bigg] \, ,
\end{align}
on the nose. From here, the considerations for the variational problem proceeds in a manner similar to the free theory. Now the action~\eqref{eq:6.2} is expected to encapsulate the entire $\alpha'$-expansion of the string effective action and the stringy version of the GHY boundary term.

Although adapting this proposal is tempting, the greatest challenge for the interacting theory appears to be constructing the maps $B_{g,n}^\mu$ themselves from first principles. We cannot directly apply the arguments of section~\ref{sec:2} to them and we must revisit the derivation of the cyclicity of string products. To this end, we first recall that the string products are defined in term of string vertices
\begin{align} \label{eq:6.4}
	\langle \mathcal{V}_{g,n+1} |  = \langle \omega | \id \otimes L_{g,n} \, ,
\end{align}
in the standard SFT formulation and string vertices are defined by integrating the string measure over regions of the appropriate moduli spaces of surfaces sufficiently away from degenerations~\cite{Zwiebach:1992ie}, according to some prescription, such as the one given by hyperbolic geometry~\cite{Costello:2019fuh,Cho:2019anu,Firat:2021ukc,Erbin:2022rgx,Firat:2023glo,Firat:2023suh,Firat:2023gfn,Firat:2024ajp,Bernardes:2024ncs}. We particularly emphasize that the string vertices that are constructed this way are graded-symmetric under the exchange of inputs
\begin{align}
	\langle \mathcal{V}_{g,n} | \Psi_1 \otimes \cdots \otimes \Psi_i \otimes \cdots \otimes \Psi_j \otimes \cdots \Psi_ n =
	(-1)^{\Psi_i \Psi_j} \langle \mathcal{V}_{g,n} | \Psi_1 \otimes \cdots \otimes \Psi_j \otimes \cdots \otimes \Psi_i \otimes \cdots \Psi_ n \, , 
\end{align}
which is demanded from the requirement of a manifestly covariant formulation.

The last property implies that the string products $L_{g,n}$ are cyclic under $\langle \omega |$ on the nose~\cite{Erler:2019loq} (i.e. $B^\mu_{g,n} = 0$ in~\eqref{eq:6.1}). This suggests that the naive relation between string vertices and string products~\eqref{eq:6.4} ignores boundary contributions and must be modified in order to account for a failure of cyclicity~\eqref{eq:6.1}. More precisely, we should use the definition
\begin{align} \label{eq:6.6}
	\langle \mathcal{V}_{g,n+1} |  = \langle \omega | \id \otimes L_{g,n}
	- \int d^D x \, \partial_\mu \langle \omega' | \id \otimes B_{g,n}^\mu \, ,
\end{align}
generally, from which the failure of cyclicity~\eqref{eq:6.1} can be trivially derived.

In order to appreciate what the maps $B_{g,n}^\mu $ are supposed to encode, we first recall that the contraction $\langle \mathcal{V}_{g,n} | \Psi^n \rangle$ takes the following form
\begin{align} \label{eq:6.7}
	\sum_{\alpha_1, \cdots, \alpha_n}
	\int {d^D k_1 \over (2\pi)^D } \cdots \int {d^D k_n \over (2\pi)^D }
	\, (2 \pi)^D \delta^{(D)}\left( \sum_{i=1}^n k_i \right) V^{(g,n)}_{\alpha_1, \dots , \alpha_n} (k_1,\cdots, k_n) 
	 \phi_{\alpha_1}(k_1) \cdots \phi_{\alpha_n}(k_n)  \, ,
\end{align}
in momentum space for the interactions between the target space fields $\phi_\alpha$~\cite{deLacroix:2017lif}. Here $V^{(g,n)}_{\alpha_1, \dots ,\alpha_n}$ encodes their off-shell interaction and contains factors that are polynomial and exponential in momenta, the latter of which roughly takes the form~$\sim e^{-c_{i j} k_i \cdot k_j}$. According to the improved definition~\eqref{eq:6.6}, we are instructed to use the momentum conservation~\eqref{eq:6.7} to eliminate the $k_1$ factors in the interactions $V^{(g,n)}$ in order to define the string products $L_{g,n}$. Upon doing so, however, we implicitly perform integration by parts and the maps $B_{g,n}^\mu$ are supposed to track these total derivatives, as we have already mentioned above.

Obviously the most important task for the interacting theory is to construct $B_{g,n}^\mu$ starting from the string vertices $\langle \mathcal{V}_{g,n} |$. However, it remains unclear how to accomplish this exactly. An initial step for the investigation could be to revisit the derivation of the $L_\infty$ relations of the string products~\cite{Erler:2019loq}. Not only must one use the improved definition~\eqref{eq:6.6} for this, but also account for the failure of the BRST identity by the boundary contributions due to~\eqref{eq:2.16}. By combining these two features, we can gain insight into the nature of $B_{g,n}^\mu$, their interrelations, and their connection to the string products. We leave a more detailed investigation of the interacting SFT for future work.

\section*{Acknowledgments}

We thank Harold Erbin, Ted Erler, Veronika Hubeny, Mukund Rangamani, and Barton Zwiebach for discussions. The work of AHF is supported by the U.S. Department of Energy, Office of Science, Office of High Energy Physics of U.S. Department of Energy under grant Contract Number DE-SC0009999 and the funds from the University of California. The work of RAM is  supported by the MIT Dean of Science Fellowship and MIT Department of Physics.

\providecommand{\href}[2]{#2}\begingroup\raggedright\endgroup

\end{document}